# Myelin figures from microbial glycolipid biosurfactant amphiphiles

Debdyuti Roy,[a] Vincent Chaleix,[c] Atul N. Parikh,[a,b,*] Niki Baccile[d,*]

[a] *Biophysics Graduate Group, University of California, Davis, Davis, California 95616, United States*

[b] *Department of Biomedical Engineering, University of California, Davis, Davis, California 95616, United States*

[c] *Université de Limoges, Faculté des sciences et techniques, Laboratoire LABCiS - UR 22722, 87060 Limoges*

[d] *Sorbonne Université, Centre National de la Recherche Scientifique, Laboratoire de Chimie de la Matière Condensée de Paris, LCMCP, F-75005 Paris, France*

*Niki Baccile, niki.baccile@sorbonne-universite.fr
* Atul Parikh, anparikh@ucdavis.edu

## Abstract

Myelin figures (MFs) – cylindrical lyotropic liquid crystalline structures consisting of concentric arrays of bilayers and aqueous media – arise from the hydration of the bulk lamellar phase of many common amphiphiles. Prior efforts have concentrated on the formation, structure, and dynamics of myelin produced by phosphatidylcholine (PC)-based amphiphiles. Here, we study the myelinization of glycolipid microbial amphiphiles, commonly addressed as biosurfactants, produced through the process of fermentation. The hydration characteristics (and phase diagrams) of these biological amphiphiles are atypical (and thus their capacity to form myelin) because unlike typical amphiphiles, their molecular structure is characterized by two hydrophilic groups (sugar, carboxylic acid) on both ends with a hydrophobic moiety in the middle. We tested three different glycolipid molecules: C18:1 sophorolipids and single-glucose C18:1 and C18:0 glucolipids, all in their nonacetylated acidic form. Neither sophorolipids (too soluble) nor C18:0 glucolipids (too insoluble) displayed myelin growth at room temperature (RT, 25 °C). The glucolipid C18:1 (G-C18:1), on the other hand, showed dense myelin growth at RT below pH 7.0. Examining their growth rates, we find that they display a linear $L \propto t$ ($L$, myelin length; $t$, time) growth rate, suggesting ballistic growth, distinctly different from the $L \propto t^{\frac{1}{2}}$ dependence, characterizing diffusive growth such as what occurs in more conventional phospholipids. These results offer some insight into lipidic mesophases arising from a previously unexplored class of amphiphiles with potential applications in the field of drug delivery.

**Introduction**

The hydration of an insoluble lamellar phase of an amphiphile gives rise to many interesting far-from equilibrium structures arising at high concentrations and at temperatures higher than their main phase transition temperature ($T_m$), at which they are in a liquid crystalline phase [1]. One class of such far-from-equilibrium structures is called myelin figures (MFs). First reported in 1854 by Virchow, MFs arise as tubular projections orthogonal to the plane of the lamellar phase of an amphiphile, typically phospholipids [2, 3]. MFs arise as a direct result of a hydration gradient created by a difference in osmotic pressure between the amphiphile-rich and aqueous phases [1] and it has previously been demonstrated that MFs grow very fast for the first 60 seconds after hydration, only for their growth to slow down significantly after a few minutes [4, 5, 6] as a result of the gradual equilibration of the hydration gradient.

A variety of biologically relevant double chain amphiphiles like 1-palmitoyl-2-oleoyl-glycero-3-phosphocholine (POPC), 1,2-dilauroyl-sn-glycero-3-phosphocholine (DLPC) and 1,2-dimyristoyl-sn-glycero-3-phosphocholine (DMPC) have been successfully used to demonstrate myelin growth [3, 7]. As a general mechanism, phospholipids self-assemble into stable lamellar bilayer stacks in water. When dry dense masses of these lamellar stacks are hydrated, the interfacial instabilities generate self-assembled finger-like projections called myelin figures, which were shown to arise from the cap edges of the bilayer stacks at the lipid-water interface and grow axially into the aqueous phase [1]. The hydration gradient causes the aqueous phase to rush into the dense bulk lamellar phase, while the amphiphiles are pushed out into the water, giving rise to a concentric array of thousands of alternating lamellae of lipid bilayers interspersed with aqueous channels. The different layers of the MFs do not grow at the same rate, with the outer tubules compressing the innermost tubules, thereby slowing down their growth rate. Eventually, as the entire lipid cake is hydrated and the difference in osmotic pressure between the bulk lipid and water phases decreases, the driving force for myelin growth ceases to exist, thereby possibly slowing the rate of myelin growth down after a few minutes [6,8].

Although these studies have provided a fundamental understanding of the mode of growth and structure of myelin figures from head-tail amphiphiles, little is known of the behavior of other classes of amphiphiles. One particular class of interest are double amphiphilic glycolipids obtained by microbial fermentation and known as microbial bioamphiphiles, or microbial biosurfactants, like sophorolipids, rhamnolipids, cellobioselipids or glucolipids [9, 10, 11]. Most of these compounds share the same feature, the presence of a mono- or disaccharide headgroup as opposed to a carboxylic acid functional group, thus providing them not only a rich and unexpected phase behavior but also a responsiveness to pH, environmental ionic strength or metal ions [12, 13, 14]. Discovered more than half a century ago [15, 16] and developed with the objective of replacing synthetic surfactants, recent studies have shown their unique self-assembly features in water, thus encouraging further work and a deeper understanding of their properties in comparison to more classical amphiphiles [12, 13]. As a matter of fact, the unique versatility in self-aggregation in water for some of these compounds [12, 13] actually questions the historical use of the term "biosurfactants", in favor of a more general "bioamphiphile" [17]. In the following, the term "(bio)surfactants" will only be used when appropriate (e.g., referring to a micellar phase), while the terms "(glyco)lipid" or "(bio)amphiphile" will be privileged most of the time (e.g., when referring to a membrane structure).

The spontaneous association of bioamphiphiles into a membrane structure, as found in vesicles for instance, is not rare [18, 19, 20]. However, it was recently shown that, at least in the case of a single-glucose lipids, their behavior resembles that of a bolaamphiphile rather than a phospholipid. For instance, the assembly process of C18:1 and C18:0 glucolipids (G-C18:1, G-C18:0) into vesicles and lamellae, respectively, is spontaneous at room temperature (RT, 25°C) in slightly acidic water; the membrane is structurally a monolayer (instead of a bilayer) and the bending rigidity is unexpectedly low (<0.5 $k_bT$) for vesicles and high (> 100 $k_bT$) for lamellae [21]. Some of these features were observed years ago in bolaamphiphilic systems [22]. Within this class of amphiphiles, only mannosylerythritole lipids (MELs), and in particular MEL-B [23], MEL-C and crude MEL mixtures [24], which are actually not bolaamphiphiles, were shown to form myelin figures as intermediate structures between the water phase and $L_\alpha$ phase in water contact experiments. However, MFs were not studied in more detail and no other microbial amphiphiles are known to form MFs.

This work explores the myelin-forming properties of three microbial bioamphiphiles, namely C18:1 sophorolipid (SL-C18:1), C18:1 glucolipid (G-C18:1) and C18:0 glucolipid (G-C18:0) (Figure 1), chosen according to their abundance and known phase behavior. SL-C18:1 is the deacetylated acidic congener of the classical sophorolipid raw mixture (acidic and lactonic), the most abundant biosurfactant referenced in literature and known to show multilamellar structures above ~60-70 wt% in water [25]. G-C18:1 can be either derived from sophorolipids through enzymatic catalysis [26] or synthesized *ex novo* using genetically modified *Candida bombicola* strains [27]. It is known to assemble into interdigitated monolayer membranes in the shape of vesicles. G-C18:0 is derived from G-C18:1 through a catalytic hydrogenation step and it is known to assemble into interdigitated monolayer in the shape of flat membranes [20, 28]. In the undertaken work, a dry cake of each of these amphiphiles was hydrated and the growth of MFs was observed using an optical microscope. When relevant, their growth rate was quantified and compared to the literature.

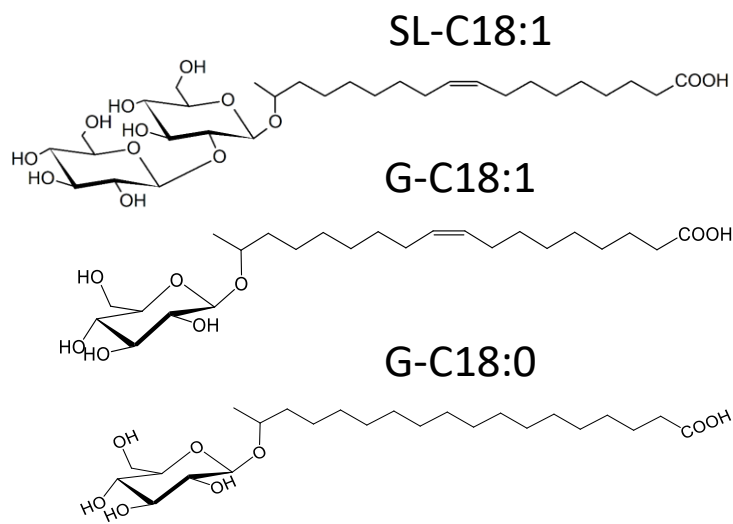

**Figure 1 – Microbial glycolipid bioamphiphile biosurfactants in their nonacetylated acidic form. Monounsaturated sophorolipid C18:1, SL-C18:1; monounsaturated glucolipid C18:1, G-C18:1; saturated glucolipid C18:0, G-C18:0. All C=C bonds are *cis* isomer form and in position 9,10.**

### Material and methods

*Chemicals.* All microbial glycolipids used in this work, monounsaturated SL-C18:1 ($M_w$ = 620 g.mol$^{-1}$), G-C18:1 ($M_w$ = 460 g.mol$^{-1}$) and saturated G-C18:0 ($M_w$ = 462 g.mol$^{-1}$), are employed from previous studies and used as such. [13, 20, 25] All molecules are nonacetylated and in their acidic form, with the C=C bond being in *cis* isomer form and in position 9,10. In terms of nomenclature, all microbial amphiphiles used here are *glycolipids*. However, more specifically, G-C18:1 and G-C18:0 only are also addressed as *glucolipids*, due to their single-glucose headgroup, differently than SL-C18:1, which contains a sophorose, di-saccharide, headgroup.

*Cleaning coverslips.* Standard Fisherbrand$^{TM}$ rectangular (50mm x 24 mm) and square (22 x 22 mm) coverslips were placed in separate glass staining dishes containing ethanol and sonicated for 5 minutes. The sonicated coverslips were then individually cleaned by dipping in chloroform and dried using a nitrogen spray gun before placing under vacuum for 10 minutes.

*Preparation of microbial glycolipid (SL-C18:1, G-C18:1, G-C18:0) solutions.* The concentration of the glycolipid solutions was 1.5 wt% (15 mg/mL) in a 70/30 v/v solvent mixture of $CHCl_3$/MeOH.

*Preparation of dried microbial glycolipid cake.* 4 μL of the glycolipid (SL-C18:1, G-C18:1, G-C18:0) solution was deposited onto the cleaned 50 mm x 24 mm rectangular glass coverslips using a glass microliter syringe. A mechanical hand press was used to seal the 22 mm x 22 mm square coverslip atop the glycolipid droplet, making a simple chip-like platform for microscopy studies. The glass support and coverslip containing the glycolipid cake was then placed in a vacuum glass desiccator for 1 hour to dehydrate the droplet thereby making a glycolipid-rich phase. Experiments are carried out at room temperature, RT, 25°C.

*Hydration of glycolipid cake.* For the study of MF growth, the dried glycolipid cakes were hydrated with water at pH 4.0, 6.0 and 8.5. A volume of 40 μL of water was pipetted in through the gap between the square and rectangular coverslips. Hydration was achieved through the capillary effect and could be observed and monitored using a brightfield optical microscope.

*Recording and quantification of myelin figure growth.* Recordings of MF growth were made using a Nikon Eclipse TE2000-S brightfield optical microscope equipped with a 20x objective (resolution is 4.5 pixel/μm). Images were captured at 20 frames per second (fps) for ~20 minutes to capture the full scale of myelin growth. If frame rates were changed during capture, the video was cropped at that time point to make two separate videos for ease of data collection and processing. The videos were analyzed using Fiji [29]. Fiji was used to measure myelin length at successive intervals of time using its measure function. Data points collected in this manner were plotted using the graphing software OriginLab to observe for trends in MFs growth rate.

**Results and discussion**

Following a standard procedure, MFs are prepared using the hydration by the contact method, which involves pipetting a small volume of water between the pressed coverslips, with water diffusing in by capillarity. The growth of myelin figures was recorded and characterized real-time using brightfield optical microscopy. Using the direct-imaging mode, the growth dynamics of the myelin figures could also be observed at millisecond time scales. All experiments were conducted at room temperature (RT, 25°C).

The first sample to be probed was the nonacetylated monounsaturated acidic sophorolipid, SL-C18:1 (Figure 1). This biosurfactant is known to assemble into a micellar phase in a broad pH and concentration range [20, 28] as well as in a lamellar structure, most likely in the shape of vesicular objects, at concentrations above 60-70 wt% [25]. Upon hydration of the SL-C18:1 cake, no myelin figures were observed. Instead, the biosurfactant mass was solubilized and washed away by the incoming waterfront. This behavior is precisely consistent with the tendency of SL-C18:1 to assemble into micelles at low concentrations and a homogeneous non-birefringent phase between 10 wt% [28, 30] and up to 60-70 wt% [25]. After several trials with similar results, the biosurfactant SL-C18:1 was not further studied for myelin growth.

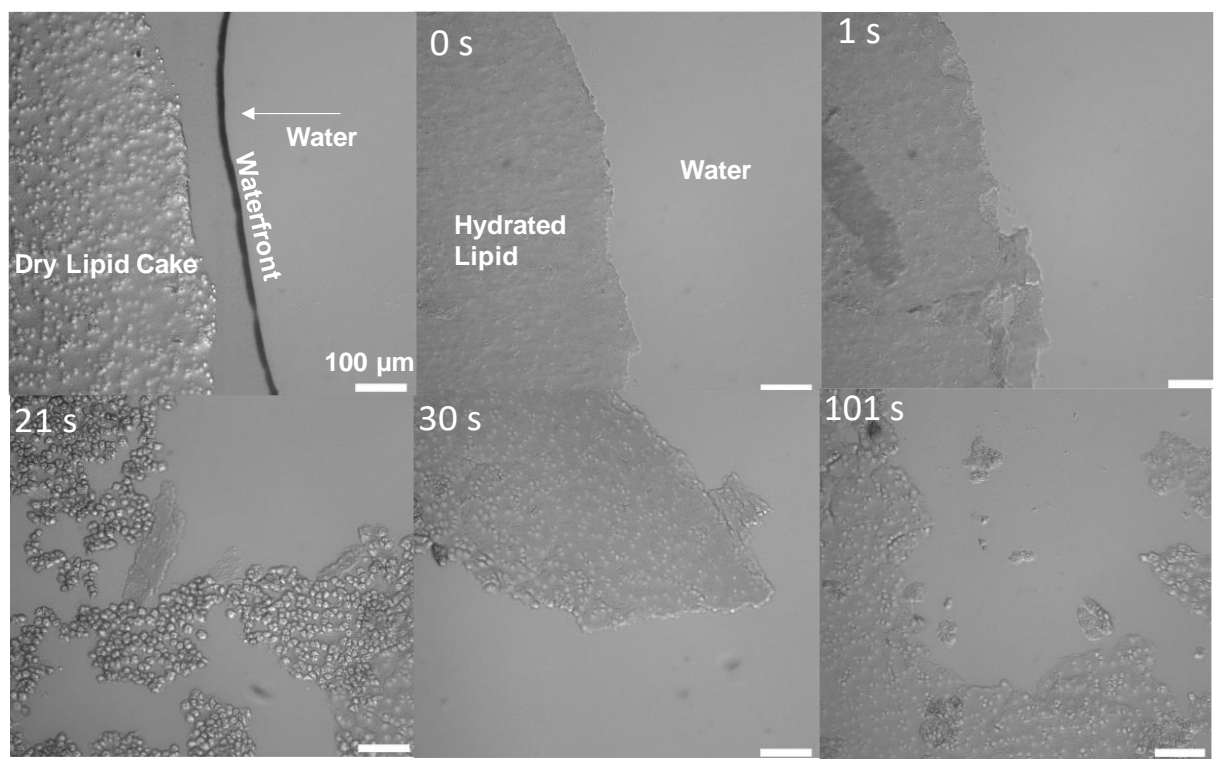

**Figure 2 - Time-resolved hydration of dried G-C18:0 cake using water at pH ~5.5 at RT. As the waterfront hits the glycolipid cake, the latter breaks up into insoluble rafts in water.**

The second molecule, glucolipid G-C18:0 (Figure 2), is known to spontaneously self-assemble into a hydrated colloidally-stable swollen lamellar phase at concentrations below 10 wt% in the pH range between 4.0 and 7.0 at RT [20, 28, 31]. The lamellae have an interdigitated structure of the size of the glucolipid's extended molecular length (about 3 nm) and are "infinitely" wide compared to the membrane thickness. The corresponding small-angle X-ray scattering (SAXS) signal is characteristics of flat membrane with a broad diffraction peak in the low wavevector range and reminiscent of swollen lamellae [20, 28, 31]. Below pH 4.0, G-C18:0 precipitates as a solid powder with lamellar crystalline structure, characterized by diffraction peaks in 1:2 wavevector ratio. Above pH 7.8, it undergoes a lamellar-to-micellar phase transition, producing water-soluble micelles. Based on these characteristics, G-C18:0 seems to be a suitable candidate for producing MFs at RT, especially below pH 7.0, in its lipid-like properties. Figure 2 shows the hydration process of the G-C18:0 dried cake, the latter visible on the left-hand side of the first image, with the waterfront in the middle and the water phase on the right-hand side. Once hydrated with water at a pH of about 5.5, G-C18:0 did not show any myelin growth. In fact, it was observed to be insoluble in water and broke apart in large raft-like fractions over time and floated away. All replicas of this experiment produced similar results. Despite the ability of G-C18:0 to assemble into colloidal lamellar structures at RT, this result is in general agreement with previous observations, which have shown that stable lamellar sheets could be obtained only after a pH jump process [20, 28] or after the supply of external energy (bath sonication, heating) to the G-C18:0 sample [31]. This sample was not studied further.

G-C18:1 (Figure 1) has a rich phase behavior, strongly dependent on the physicochemical conditions of the aqueous medium [13]. In bulk water at a pH below 3.5, it precipitates as a crystalline powder with lamellar structure. Between pH values of about 3.5 and 6.2, it spontaneously forms stable vesicular objects with curiously low bending rigidity, while a vesicle-to-micelle phase transition starts above pH 6.2 until about 7.0 [20, 28]. Within the context of MFs, G-C18:1 was studied below and above pH 7.0, more specifically at pH 8.5 (micellar phase), pH 4.0 and 6.0 (vesicle phase). However, at pH 8.5, MFs are not observed or are not stable, as G-C18:1 is essentially water soluble. This result is actually in agreement with what it is observed for acidic SL-C18:1 sophorolipids. If MFs were instead systematically observed at both pH 4.0 and pH 6.0, it was quite challenging to find a region of interest of the cake where MFs grew homogeneously in time, as flooding was often observed, as discussed further below.

Time-resolved microscopy images (Figure 3a, Figure S 1) and the corresponding Videos S1-S4 in the Supporting Information show the details of the MFs observed for G-C18:1 at both pH 4.0 and pH 6.0. As the waterfront reaches and it begins to hydrate the desiccated cake, finger-like projections begin to appear almost instantaneously at the glucolipid-water interface (Figure 3a,b) and then slow down considerably within 60 s. MFs' growth was observed and followed as a front, rather than as single myelin figure. These observations, as well as the shape of individual myelin figures (Figure S 2), agree well with the characteristics of MF growth seen in systems using POPC, DOPC and DMPS (1,2-dimyristoyl-sn-glycero-3-phospho-L-serine)

[3]. The MFs produced by G-C18:1 were also structurally similar to those observed in PC-based MFs, with a coiling, cylindrical shape (Figure 3a, b).

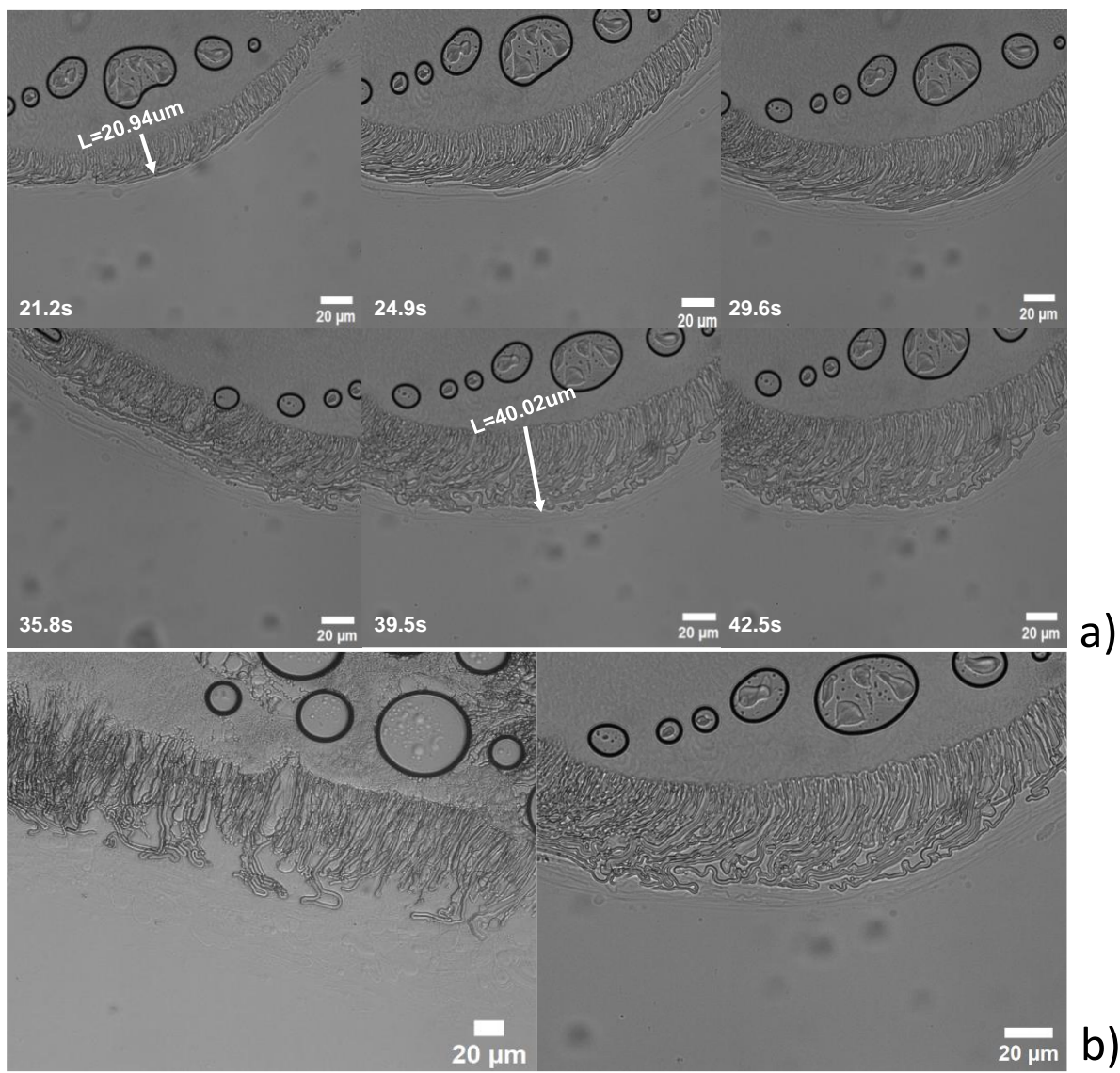

**Figure 3 – Growth of MFs from G-C18:1 at RT. a) Time-resolved evolution of myelin figures from using water at pH 4.0; b) Typical myelin figures observed in two different samples prepared from water at pH 6 (left-hand image) and pH 4 (right-hand image).**

However, deviating from PC-based myelin growth trends, most G-C18:1 samples displayed initial myelin growth occurring at a higher than usual rate, followed by consequent dissociation from the bulk glucolipid phase, as seen in various moments in Video S3 (e.g., time stamp: 15-50 s). This was a frequently observed artifact which did not allow for a proper follow-up of the growth. The origin of this was attributed to an improperly assembled chip combined with the more hydrophilic nature of G-C18:1 compared to PC-based amphiphiles. Ideally, a properly made chip will have a fully desiccated cake with minimal distance between the two coverslips.

If the spacing between the coverslips is larger than permissible, water can diffuse in from above, affecting both faces of the cake, thus flooding the sample. In this case, myelins detach from the lipid cake owing to the flow of water. On the other hand, even when this phenomenon does not occur, the G-C18:1 cake hydrates rapidly, causing water flooding in the weakest regions of the cake. When following the growth of the MFs over time in this specific case, unrealistic growth rates of MF length (*L*) vs. time (*t*) were observed (Figure 4, sample *f\**, Video S3). Typically, the *L*(*t*) plot measured at the forefront of the MFs shows a $L \propto t^\alpha$, with $\alpha$= 3.17 ± 0.06 μm/s (Table 1), in obvious disagreement with all other samples, as discussed below.

In the absence of artifacts, growth rate of MFs was estimated from Videos S1, S2, and S4, each collected on different replicas. The rate of growth was calculated by measuring the myelin length from their growth front to their base at different time points, presented as *L*(*t*) plot in Figure 4. For most sample sets, except one, the earliest time point was captured above about 10-15 s after hydration and the longest at ~100 s. The fact that we could hardly measure time points below 10 s is explained by the time required to find a reliable region of the cake where flooding did not affect the measurement. After many trials, only one attempt, clearly labeled as *initial growth* in Figure 4, was retained.

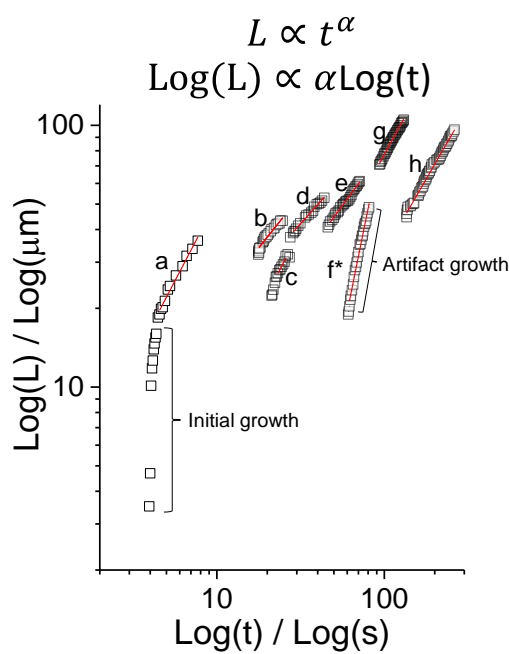

**Figure 4 – Myelin length (*L*) vs time (*t*) plot for myelin figure growth prepared from G-C18:1. Samples were hydrated at RT (25°C) using milliQ-grade water at pH 4 (datasets *c*, *d*, *e*, *f\**, *g*) and pH 6 (datasets *a*, *b* & *h*). The data are extracted from live Video S1 through S4 given in the Supporting Information. Table 1 associates each dataset (*a* through *h*) to the given video. Dataset *f* is marked with \* to indicate that the growth rate measured for this sample is affected by an artefact (flooding of myelins, discussed in main text).**

Overall, Figure 4 indicates that MFs show a linear growth rate (log-log scale) for all data sets, belonging to three different replicas. Datasets *c*, *d*, *e* and *g* all belong to the same sample and were recorded at different locations of the hydrated cake, thus explaining the shift in time. All data sets were extracted from videos S1-S4 shown in the Supporting Information and are presented in the table shown in Figure 4. The growth rate of MFs few seconds after hydration follows a direct proportionality between time and myelin length, from which the rate can be obtained directly from the *L*(*t*) data in log-log scale. Excluding sample *f\** (flooding of myelins), the average growth rate, essentially linear, with α= 1.02 ± 0.23 μm/s (Table 1). The error corresponds to the standard deviation of α reported for datasets *a* through *h* (excluding *f\**), while the error for α corresponds to the uncertainty associated to the corresponding linear fitting process in the *L*(*t*) plots (log-log scale) in Figure 4.

**Table 1** – **Values of the growth rate exponent α (μm/s) for datasets *a* through *h* and related pH of the wetting water phase. The source video (Supporting Information) is associated to a given dataset, which corresponds to eight different locations of myelin growth found in four replicas. The error associated to α for each datasets is obtained from the linear regression in the myelin length (*L*) vs time (*t*) plot (log-log scale) presented in Figure 4. The error associated to the Average value of α is the standard deviation of datasets *a* through *h*, excluding dataset *f\**, the latter being characterized by flooding of the myelins.**

| Sample | pH | α / μm/s |
|---|---|---|
| *a* (Video S1a) | 6.0 | 1.22 ± 0.05 |
| *b* (Video S1b) | 6.0 | 0.80 ± 0.05 |
| *c* (Video S2) | 4.0 | 1.28 ± 0.11 |
| *d* (Video S2) | 4.0 | 0.71 ± 0.02 |
| *e* (Video S2) | 4.0 | 0.87 ± 0.02 |
| *f\** (Video S3) | 4.0 | 3.17 ± 0.06 |
| *g* (Video S2) | 4.0 | 1.17 ± 0.01 |
| *h* (Video S4) | 6.0 | 1.11 ± 0.01 |
| | | |
| Average (*a/b/c/d/e/g/h*) | | 1.02 ± 0.23 |

In previously conducted experiments using PC-based amphiphiles, length of MFs displays the explosive beginning and consequent slowing down of MFs growth over time to vary as a function of $\sqrt{t}$ [4, 32, 33]. MFs growing from samples of G-C18:1 also show such a double rate of growth. However, rate growth rate above few seconds from hydration deviate from $L \propto \sqrt{t}$ by showing a linear growth rate. This observed $L \propto t$ growth trend is curious, and suggestive of a rapid, non-diffusive growth of myelins. Such a ballistic growth profile is consistent with the motion of a particle in Brownian dynamics at short time scales [34]. Under these conditions, the collision-free movement of the particle accelerates the dynamics. In our case, the strong hydration gradient, we surmise, produces a correspondingly strong water permeation flux inducing local stresses in the lamellar phase. This stress then drives large scale collective undulations, which amplify. The resulting directed growth of individual myelins into the aqueous phase then occurs through swift, unswerving, collective undulations of lamellar membranes into the aqueous phase akin to the ballistic motion.

As the entire G-C18:1 cake was hydrated, MF growth rate was observed to slow down considerably, usually stalling completely at the 10-minute mark. There are two broad schools of thought, which attempt to explain the growth of MFs. The first mechanistic model proposes that myelin figures grow through a collective diffusive process ultimately producing semi-stable structures of the lowest free energy in local equilibrium with the surrounding water [1, 7, 35]. The second model postulates that they are dynamic structures formed transiently as a result of a driving stress, which in the present case is the hydration gradient at the amphiphile-water interface [8]. The latter school of thought appears to be a better representation of our observations. The ballistic and almost instantaneous growth of MFs observed in the first 60 s could be a result of the strongest hydration gradient between the bulk water and the glucolipid-rich phases. It also suggests that the growth kinetics are neither controlled by the collective diffusion of the glucolipid nor by its swelling in the dry state because of the water influx: a more complex dynamic self-organization must be at works. As the entire glucolipid phase gets hydrated and the hydration gradient decays, ultimately ceasing to exist, it causes MF growth to slow down and eventually stall. However, the MFs tend to remain in their stalled state instead of returning to the lamellar bilayer once there is no longer a driving force. This is most likely because the myelin figures are in a trapped kinetic or metastable state rather than a locally equilibrated one.

**Conclusions**

This work addresses the myelin formation of selected microbial biosurfactant bioamphiphiles, nonacetylated acidic sophorolipids (C18:1) and glucolipids (C18:1, C18:0). Most of the previous work in this field deals with phosphatidylcholine-based amphiphiles, like POPC and DOPC, which have zwitterionic headgroups and 2-chain hydrophilic tails. Experiments involving this class of amphiphiles have shown that MFs are supramolecular liquid crystalline organizations of amphiphiles which show diffusion-based growth on application of a hydration gradient, and grow from the edges of the lamellar stacks, with the gradient adding in amphiphiles from the root of the MF. The structure can be described as a cylindrical, concentric array of amphiphiles interspersed with layers of aqueous medium.

G-C18:0 has a saturated stearic acid tail, while G-C18:1 has a monounsaturated (C9,10) oleic acid tail, similarly to sophorolipids. At RT, SL-C18:1 is solubilized, most likely due to its water solubility (micellar phase) at low concentrations, while G-C18:0 was, on the contrary, insoluble and did not show any myelin growth. On the other hand, G-C18:1 showed dense myelin growth at pH below 7 and no growth at pH 8.5, as at this pH, G-C18:1 mainly forms a micellar phase. Similar to phospholipids, MFs from G-C18:1 were observed to grow at two rates, an explosive one within the first few seconds from wetting and a slower one, which was observed to be linear with respect to time, thus showing a deviation from the $L \: \alpha \: t^{\frac{1}{2}}$ dependency seen in previous studies. Myelin growth eventually stops within the first 10-15 minutes. Beyond this point, the MFs maintained their structure and neither grew not retracted back into the bulk glucolipid phase. This points to MF growth being representative of a non-equilibrium non-steady state, and being dependent on an external driving force, in this case the hydration gradient, removal of which causes the MFs to attain a non-equilibrium steady state.

**Acknowledgements**

The France-USA Fulbright Commission (award N° PS00342290) is kindly acknowledged for supporting this research. Additional support for the work at UC Davis was obtained through a grant from the National Science Foundation (Award # 2104123). Daniel Speer (University of California, Davis) is kindly acknowledged for his assistance in preparing the myelins figures.

**Supporting Information available**

*Figures*

**Figure S 1 - Time-resolved evolution of myelin figures from G-C18:1 using water at pH 6.0 at RT (25°C). Images are extracted from Video S4.**

**Figure S 2 – Close-up image of individual myelins obtained from G-C18:1 using a water bath at pH 4.0 at RT (25°C)**

*Multimedia*

**Video S 1 – MFs growth for G-C18:1 at RT as a function of time. a) Video recorded upon hydration during the first few seconds. *b*) growth occurring after about 15 s of hydration.**

**Video S 2 - MFs growth for G-C18:1 at RT as a function of time recorded on a newly-**

**prepared sample**

**Video S 3 -** MFs growth for G-C18:1 at RT as a function of time recorded on a newly-prepared sample. Effect of flooding showed here

**Video S 4 -** MFs growth for G-C18:1 at RT as a function of time recorded on a newly-prepared sample.

# Supplementary Information

# Myelin figures from microbial glycolipid biosurfactant amphiphiles

Debdyuti Roy[a], Vincent Chaleix[c], Atul Parikh,[a,b,*], Niki Baccile[d,*]

[a] *Biophysics Graduate Group, University of California, Davis, Davis, California 95616, United States*

[b]*Department of Biomedical Engineering, University of California, Davis, Davis, California 95616, United States*

[c]*Université de Limoges, Faculté des sciences et techniques, Laboratoire LABCiS - UR 22722, 87060 Limoges*

[d] *Sorbonne Université, Centre National de la Recherche Scientifique, Laboratoire de Chimie de la Matière Condensée de Paris, LCMCP, F-75005 Paris, France*

*Niki Baccile, niki.baccile@sorbonne-universite.fr
* Atul Parikh, anparikh@ucdavis.edu

**Multimedia supports**

**Video S 1 – MFs growth for G-C18:1 at RT as a function of time. a) Video recorded upon hydration during the first few seconds. *b*) growth occurring after about 15 s of hydration.**

**Video S 2 - MFs growth for G-C18:1 at RT as a function of time recorded on a newly-prepared sample**

**Video S 3 - MFs growth for G-C18:1 at RT as a function of time recorded on a newly-prepared sample. Effect of flooding showed here**

**Video S 4 - MFs growth for G-C18:1 at RT as a function of time recorded on a newly-prepared sample.**

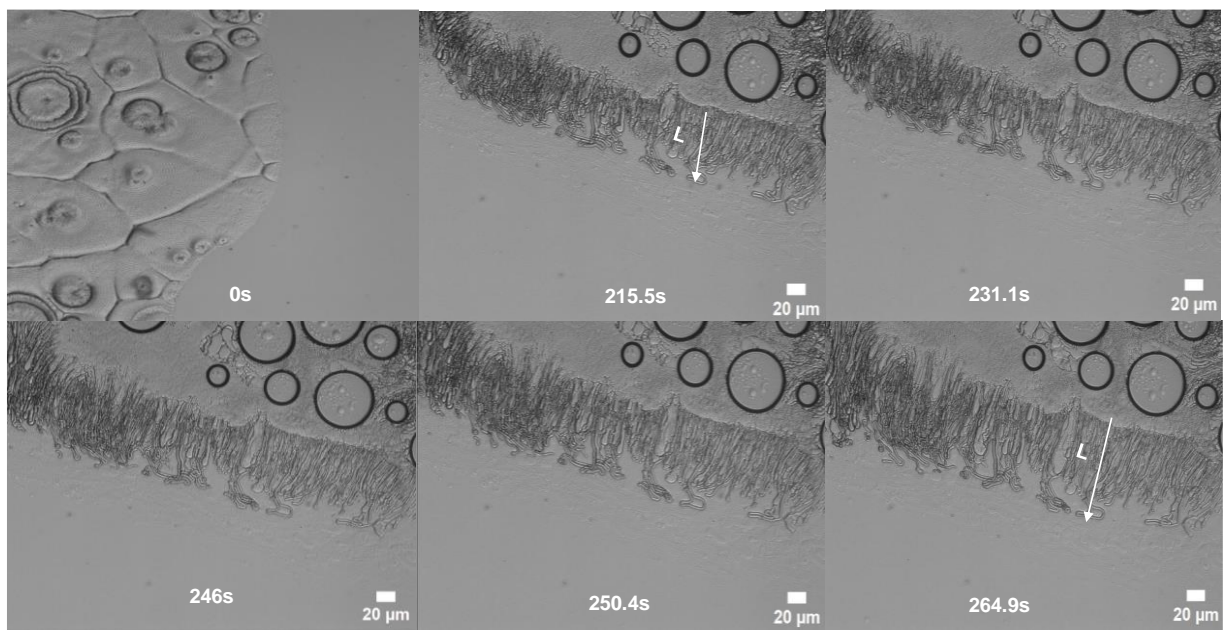

**Figure S 1 - Time-resolved evolution of myelin figures from G-C18:1 using water at pH 6.0 and RT. Images are extracted from Video S4.**

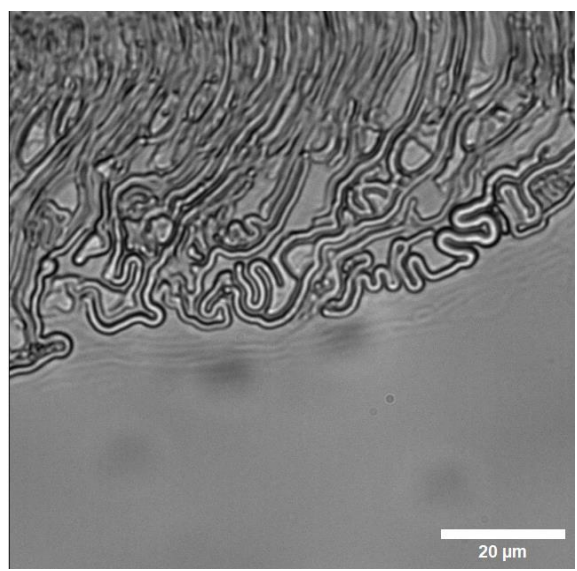

**Figure S 2 – Close-up image of individual myelins obtained from G-C18:1 using a water at pH 4.0 at RT (25°C)**

2